\documentclass[12pt]{iopart}
\usepackage[T1]{fontenc}
\usepackage[latin9]{inputenc}
\usepackage{color}
\expandafter\let\csname equation*\endcsname\relax
\expandafter\let\csname endequation*\endcsname\relax
\usepackage{amsmath}
\usepackage{amssymb}
\usepackage{graphicx}
\usepackage[normalem]{ulem}

\makeatletter

\providecommand{\tabularnewline}{\\}

\@ifundefined{textcolor}{}
{%
 \definecolor{BLACK}{gray}{0}
 \definecolor{WHITE}{gray}{1}
 \definecolor{RED}{rgb}{1,0,0}
 \definecolor{GREEN}{rgb}{0,1,0}
 \definecolor{BLUE}{rgb}{0,0,1}
 \definecolor{CYAN}{cmyk}{1,0,0,0}
 \definecolor{MAGENTA}{cmyk}{0,1,0,0}
 \definecolor{YELLOW}{cmyk}{0,0,1,0}
}

\@ifundefined{definecolor}
 {\usepackage{color}}{}
\@ifundefined{definecolor}
 {\usepackage{color}}{}
\usepackage{stmaryrd}

\def\kbar{{\mathchar'26\mkern-9mu k}}
\def\pred{\mathfrak{p}}

\makeatother

\makeatother

\begin{document}

\title{Phase diagram of the Anderson transition with atomic matter waves}

\author{M.~Lopez$^1$, J.-F.~Cl\'ement$^1$, G.~Lemari\'e$^{2,3}$, D.~Delande$^3$, P.~Szriftgiser$^1$, J.~C.~Garreau$^{1,4}$}

\address{$^1$Laboratoire de Physique des Lasers, Atomes et Mol\'ecules, Universit\'e
Lille 1 Sciences et Technologies, CNRS; F-59655 Villeneuve d'Ascq
Cedex, France}

\address{$^2$Laboratoire de Physique Th\'eorique UMR-5152, CNRS and
  Universit\'e de Toulouse, F-31062 France}

\address{$^3$Laboratoire Kastler-Brossel, UPMC-Paris 6, ENS, CNRS; 4 Place Jussieu,
F-75005 Paris, France}

\address{$^4$Corresponding author}
\ead{jean-claude.garreau@univ-lille1.fr}

\date{\today}
\begin{abstract}
We realize experimentally a cold atom system equivalent to the 3D Anderson
model of disordered solids where the anisotropy can be controlled 
by adjusting an experimentally accessible parameter.
This allows us to study experimentally the disorder vs anisotropy phase diagram of the Anderson metal-insulator transition.
Numerical and experimental data compare very well with
each other and a theoretical analysis based on the self-consistent theory
of localization correctly discribes the observed behavior,
illustrating the flexibility of cold atom experiments
for the study of transport phenomena in complex quantum systems.
\end{abstract}


\maketitle

\section{Introduction}

The interplay of disorder and quantum interference has been an important
subject in physics for more than 50 years. Quantum interferences,
which are at the heart of most quantum effects, rely on precise relative
phases between quantum trajectories, which are strongly sensitive
to perturbations like decoherence (i.e. coupling with a large environment)
and scattering of the wave function in potential wells. This last
effect becomes particularly difficult to describe theoretically in
a disordered system, in which these scattering processes have a random
character. Intuitively, one easily understands that such kind of effect
shall play an important role for example in the low-temperature electric
conductance of solids. In fact, Anderson showed in 1958 that the presence
of disorder might produce a \emph{spatial localization} of the wavefunction,
which \emph{suppresses} conductivity \cite{Anderson:LocAnderson:PR58}
thus the name of ``strong'' localization.

Laser cooling opened the possibility of realizing very clean experiments
in disordered systems, which generated a burst of interest on the
subject. In adequate conditions, ultracold atoms placed in spatially
structured laser beams feel this radiation as a mechanical potential
acting on the center of mass variables of the atoms. Disordered potentials
created in such a way allowed the realization of spatially disordered
systems in one dimension \cite{Billy:AndersonBEC1D:N08,Roati:AubryAndreBEC1D:N08}
and three dimensions \cite{Kondov:ThreeDimensionalAnderson:S11,Jendrzejewski:AndersonLoc3D:NP12}. 
Despite
these progresses, the Anderson metal-insulator transition (which manifests
itself in 3 or more dimensions) is still very difficult to study in
such systems, because Anderson
localization requires a very strong disorder and  -- the cold atomic samples
being prepared in the absence of disorder -- the energy distribution 
of the atoms unavoidably spreads across  the so-called \emph{mobility edge},
an energy threshold separating localized and extended eigenstates.
This in turn implies that the localized fraction, which can be \emph{directly} measured in cold-atom experiments
from the temporal evolution of the spatial probability distributions, remains small.
Fortunately,
one can find other systems also described by the Anderson localization physics,
which are not a direct transposition of the condensed matter system, 
 but rely on the profound analogy between
quantum chaotic systems and disordered systems \cite{Efetov:SupersymmetryInDisorder:97}. 
Using the quasiperiodic kicked rotor (QpKR) \cite{Casati:IncommFreqsQKR:PRL89}, an effectively 3D variant
of the paradigmatic system of quantum chaos \cite{Casati:LocDynFirst:LNP79},
the Anderson transition has been observed, its critical
exponent measured experimentally \cite{Chabe:Anderson:PRL08,Lemarie:AndersonLong:PRA09},
its critical wavefunction characterized \cite{Lemarie:CriticalStateAndersonTransition:PRL10},
and its class of universality firmly established \cite{Lopez:ExperimentalTestOfUniversality:PRL12},
making this system an almost ideal environment
to study Anderson type quantum phase transitions. 

One advantage of this cold atom chaotic system as compared to other disordered systems is that the disorder can be controlled 
very precisely: the mean free path and the anisotropy are two experimentally tunable parameters.
This allows us to present in this work an experimental
study of the disorder vs anisotropy phase diagram of the Anderson transition,
as well as an analytical description of these properties based on
the self-consistent theory of Anderson localization, which brings
another important brick to our detailed knowledge on the Anderson metal-insulator transition.

\section{Controlled disorder and anisotropy within a cold atom system}

The quasiperiodic kicked rotor is described by the one-dimensional time dependent
Hamiltonian
\begin{equation}
H_{\mathrm{qpkr}}=\frac{p^{2}}{2}+K\cos x\left(1+\varepsilon\ \cos\omega_{2}t\ \cos\omega_{3}t\right)\sum_{n}\delta(t-n).\label{eq:Hqpkr}
\end{equation}
Experimentally, it is realized by placing laser-cooled atoms (of mass
$M$) in a standing wave (formed by counterpropagating beams of intensity
$I_{0}$ and wavenumber $k_{L}$) which generates an effective sinusoidal mechanical
potential -- nicknamed ``optical potential'' --  $\cos x$ acting on the center of mass position $x$
of the atom. The
standing wave is modulated by an acousto-optical modulator in order
to form a periodic (of period $T_{1}$) train of short square pulses
whose duration $\tau$ is short enough that, at the time scale of
atom center of mass dynamics, they can be assimilated to Dirac $\delta$-
functions. The amplitude of such pulses is further modulated with frequencies
$\omega_{2}$ and $\omega_{3}$, proportionally to $1+\varepsilon\ \cos\omega_{2}t\ \cos\omega_{3}t$. 
Lengths are measured in units of
$\left(2k_{L}\right)^{-1}$, time in units of $T_{1}$, momenta in
units of $M/2k_{L}T_{1}$; note that $\left[x,p\right]=i\kbar$ with
$\kbar=4\hbar k_{L}^{2}T_{1}/M$ playing the role of a reduced Planck
constant. The pulse amplitude is $K=\kbar\tau\Omega^{2}/8\Delta_{L}$,
where $\Omega$ is the resonance Rabi frequency between the atom and
the laser light and $\Delta_{L}$
the laser-atom detuning. Fixed parameters used throughout the present work are $\kbar=2.885$,
$\omega_{2}=2\pi\sqrt{5}$, $\omega_{3}=2\pi\sqrt{13}$ \footnote{Rational values of 
$\omega_2/2\pi, \omega_3/2\pi$ produce a periodically -- instead of quasiperiodically -- kicked
rotor, with different long time behaviour\cite{Ringot:Bicolor:PRL00,Lignier:SubFMecs:EPL05,Lignier:Reversibility:PRL05}. We chose ``maximally
irrational ratios'' to avoid this problem.}.

If $\varepsilon=0$ one obtains the standard
kicked rotor, which is known to display fully chaotic classical dynamics
for $K\ge6$ \cite{Chirikov:ChaosClassKR:PhysRep79}. 
At long time, the dynamics is a so-called chaotic diffusion in momentum space,
which is -- although perfectly deterministic -- characterized by a diffusive increase of the r.m.s. momentum:
$\langle p(t) - p(t=0) \rangle=0,\ \langle [p(t)-p(0)]^2\rangle \sim 2Dt$ (which $D$ the
classical diffusion constant) where the average $\langle \rangle$ is performed  
over an ensemble of trajectories associated with neighbouring  initial conditions.
The statistical distribution of $p(t)$ has the characteristic Gaussian shape of a diffusion process,
whose width increases like $\sqrt{t}.$
Quantum mechanically,
this system displays the phenomenon of \emph{dynamical localization},
that is, an asymptotic saturation of the average square momentum $\left\langle p^{2}\right\rangle $
\cite{Casati:LocDynFirst:LNP79} at long time, that is localization in momentum space
instead of chaotic diffusion, which has been proved to be a direct
analog of the Anderson localization in one dimension \cite{Fishman:LocDynAnders:PRL82, Altland:TheoryKR:PRL96, Tian:TheoryKR:NJP10}.

If $2\pi/T_{1},\omega_{2},\omega_{3},\kbar$ are incommensurable and
$\varepsilon\neq0$ one obtains the QpKR, which can be proven to be equivalent to the Anderson model in 3 dimensions
\cite{Casati:IncommFreqsQKR:PRL89, Lemarie:AndersonLong:PRA09, Lemarie:These:09, Tian:TheoryAndersonTransition:PRL11}.
In a nutshell, the QpKR which is a 1-dimensional system with a time-dependent Hamiltonian
depending on 3 different quasi-periods, can be mapped on a kicked ``pseudo''-rotor, a 3-dimensional system with a time periodic
Hamiltonian. As shown in detail in \cite{Lemarie:AndersonLong:PRA09}, both systems share the same temporal dynamics. The Hamiltonian of the
pseudo-rotor is:
\begin{equation}
\mathcal{H}=\frac{p_{1}^{2}}{2}+\omega_{2}p_{2}+\omega_{3}p_{3}+K\cos x_{1}\left[1+\varepsilon\cos x_{2}\cos x_{3}\right]\sum_{n}\delta(t-n)\;,
 \label{eqKR3DquasiperH}
 \end{equation}
with an initial condition taken as a planar source in momentum space (completely delocalized along the transverse directions $p_2$ and $p_3$).
Note that the kinetic energy has a different dependence on the momentum in each direction:
standard (quadratic) in direction 1, but linear in directions 2 and 3; hence, the name pseudo-rotor. 

The Hamiltonian (\ref{eqKR3DquasiperH}) is periodic in configuration space. It can thus
be expanded in a discrete momentum basis composed of states $\vert \boldsymbol \pred \rangle = \vert \boldsymbol p = \kbar \boldsymbol \pred\rangle,$
where the $\pred_{i}$ are integers \footnote{This implies periodic boundary conditions. In general -- especially
for an ''unfolded`` rotor for $x_1$ is a position in real space as realized in the experiment -- one should
use the Bloch theorem which guarantees the existence of states whose wavefunction take a phase factor after
translation by $2\pi.$ This amounts at considering not integer $\pred_{i}$ values, but rather
$\pred_{i}=n_i+\beta_i$ with $n_i$ an integer and $\beta_i$ a fixed quantity called quasimomentum. All conclusions
obtained in the simplest case $\beta_i=0$ can be straightforwardly extended in the general case.}. In this basis, 
the evolution operator over one temporal period writes as the product $U = J V$ of
an on-site operator: $ V(\boldsymbol{\pred})= e^{-i \phi_{\boldsymbol{\pred}}}$ with phases $\phi_{\boldsymbol{\pred}} = \frac{\kbar \pred_{1}^{2}}{2}+\omega_{2}\pred_{2}+\omega_{3}\pred_{3}$
and of a hopping operator $J$ such that:
\begin{equation}
 \langle \boldsymbol \pred_f \vert J \vert \boldsymbol \pred_i \rangle = \int \frac{d \boldsymbol x }{(2 \pi)^3} 
\exp{\left[-i \frac{K\cos x_{1}\left(1+\varepsilon\cos x_{2}\cos x_{3}\right)}{\kbar}\right]} 
\ \exp{\left[- i (\boldsymbol \pred_i -\boldsymbol \pred_f)   \boldsymbol x\right]} \; .
\end{equation}
The phases $ V(\boldsymbol{\pred})$ are different on each site of the momentum lattice, and, although perfectly
deterministic, constitute a pseudo-random sequence completely analogous to the true random on-site
energies of the Anderson model. This makes it possible to identify $V$ as the disorder operator for the QpKR.
The parameter $K$ controls the hopping amplitudes, that is the transport properties
in the absence of disorder. The larger $K$, the larger distance the system propagates in momentum space (with the operator $J$)
before being scattered by the disorder operator $V$. As shown below, the associated mean free path in momentum space
is of the order of $K/\kbar.$ Rather counter-intuitively, the weak disorder limit
then corresponds to the large $K$ limit, that is strong pulses, while the strong disorder
limit where Anderson localization is expected corresponds to small $K.$ It should also be stressed that, for very small $K,$
(very strong disorder), the system remains frozen close to its initial state, with a trivial on-site Anderson
localization. This is not really surprizing as the classical dynamics is then regular instead of chaotic
and even the classical chaotic diffusion is suppressed.

In the following, we will concentrate on the role of the $\varepsilon$  parameter, which drives the anisotropy between the transverse directions and the longitudinal direction, showing the analogy of \eqref{eqKR3DquasiperH} 
with a system of weakly coupled disordered chains as considered in \cite{Soukoulis:AnisotropicAnderson:PRL96}.

With such a system, we experimentally observed and characterized the
Anderson transition \cite{Chabe:Anderson:PRL08, Lemarie:AndersonLong:PRA09},
which manifests itself by the fact that the momentum distribution is
exponentially localized $\Pi(\pred;t)\sim\exp\left(-\left|\pred\right|/\pred_{\text{loc}}\right)$ (with $\pred_{\text{loc}}$
the localization length) if $K$ is smaller than a critical value $K_{c}(\varepsilon)$ and
Gaussian diffusive $\Pi(\pred;t)\sim\exp\left(-\pred^{2}/4 D t\right)$ (where $D$ is the diffusion coefficient)
for $K>K_{c}(\varepsilon)$ after a sufficiently long time. 
At criticality, $K=K_c(\varepsilon)$, the localization length diverges, the diffusion constant vanishes,
and the critical state is found \cite{Lemarie:CriticalStateAndersonTransition:PRL10} to have a characteristic 
Airy shape
\begin{equation}
\label{eq:Airy}
\Pi(\pred;t)\approx \frac{3}{2}\frac{ \alpha }{ \sqrt{\Lambda_c(\varepsilon) t^{2/3}}}  \text{Ai}\left[\alpha \sqrt{ \frac{\vert \pred\vert^2}{\Lambda_c(\varepsilon) t^{2/3}}}\right]
\end{equation}
following the anomalous diffusion at criticality:
$\langle \pred^{2}\rangle = \Lambda_{c}(\varepsilon) t^{2/3}$ (here $\alpha=3^{1/6} \Gamma(2/3)^{-1/2}$). 
The fundamental quantities characterizing the threshold 
of the transition are therefore $K_c(\varepsilon)$ and $\Lambda_c(\varepsilon)$, and we will consider in the following their dependence as a function of the anisotropy parameter $\varepsilon$.

\section{Experimental determination of the anisotropy phase diagram}

\begin{figure}
\begin{centering}
\includegraphics[clip,width=0.8\linewidth]{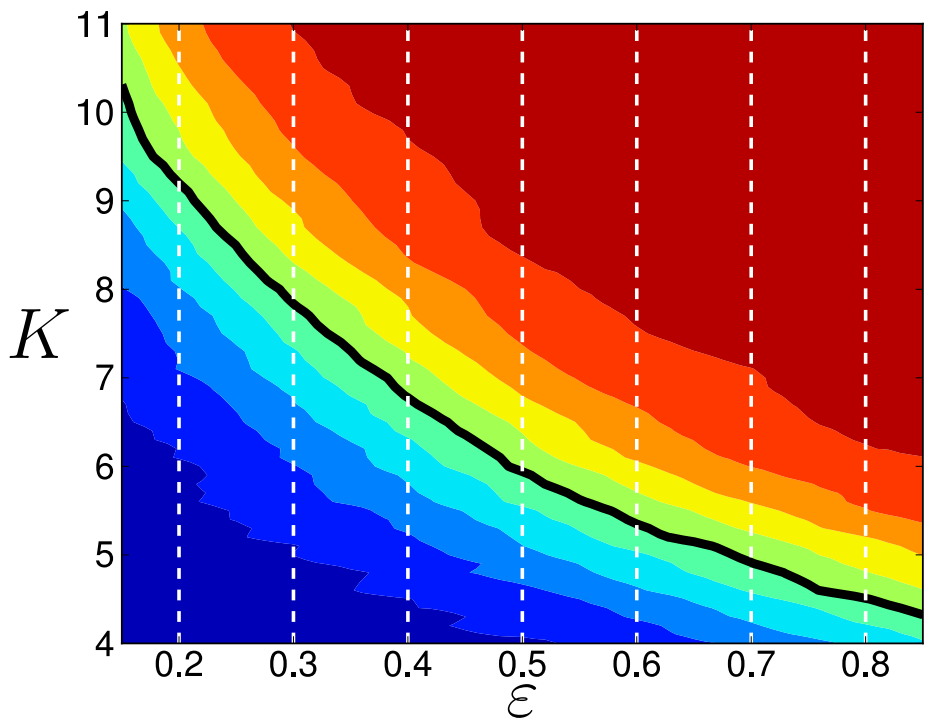}
\par\end{centering}

\caption{\label{fig:Paths}Schematic phase diagram of the metal-insulator 
Anderson transition for the quasi-periodic kicked rotor. The color plot corresponds to
growth rate $\alpha$ of $\langle p^2(t)\rangle\propto t^{\alpha}$ at long time
(1000 kicks for this plot), estimated from numerical simulations. Blue color represents localization ($\alpha=0$),
red represents diffusive dynamics ($\alpha=1$). The black line corresponds to $\alpha=2/3,$ that
is the critical line of the Anderson transition. Paths (white dashed lines) form the grid used
for the determination of $K_{c}(\varepsilon)$.}
\end{figure}

\begin{figure}
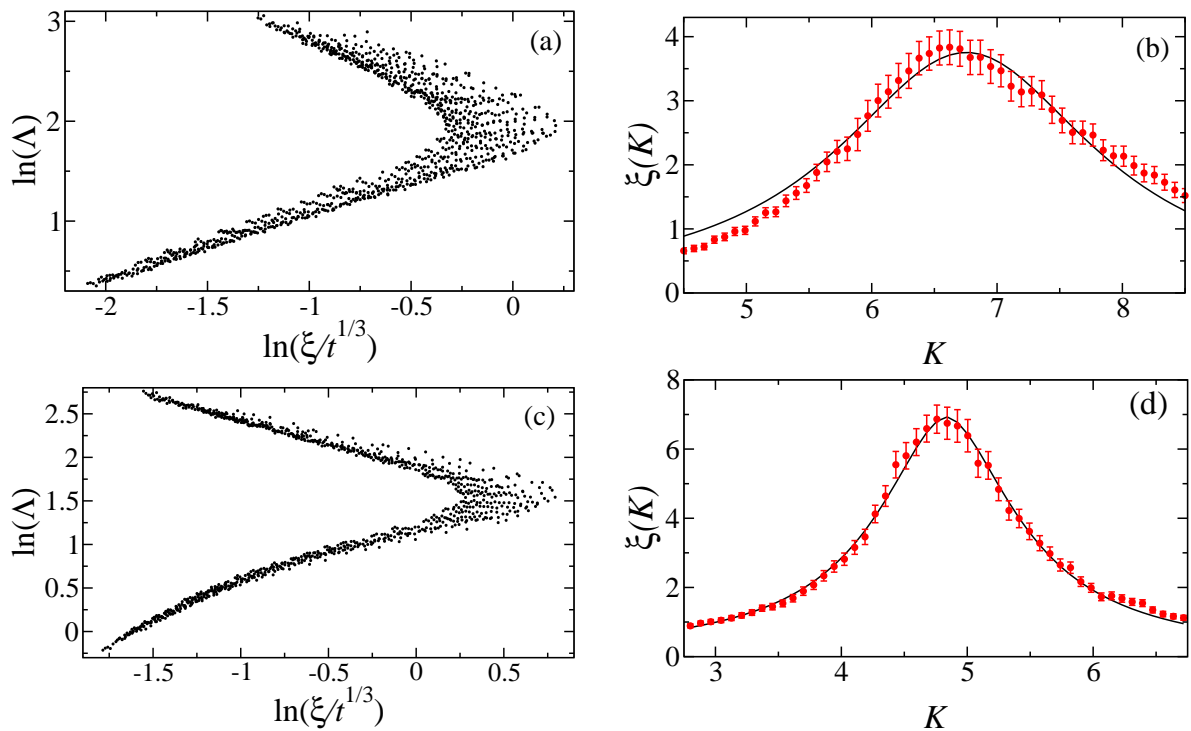

\begin{centering}
\includegraphics[width=0.48\linewidth]{figure2a.eps}\hskip 0.04\linewidth \includegraphics[width=0.48\linewidth]{figure2b.eps}\\
\includegraphics[width=0.48\linewidth]{figure2c.eps}\hskip 0.04\linewidth \includegraphics[width=0.48\linewidth]{figure2d.eps}
\par\end{centering}

\caption{\label{fig:Scaling}Determination of the critical point from experimental
data at two different anisotropy $\varepsilon$. The finite-time scaling method (see text)
applied to the experimental data $\Lambda(t) \propto \Pi(\pred=0;t)^{-2} t^{-2/3}$ allows for a determination of the
scaling function $F$ (Eq.\ \eqref{eq:scafunc}) represented in a and c and the scaling parameter $\xi(K)$ shown in b and d. The critical
point corresponds to the tip at the right of the scaling function (see a and c),
at the intersection of the diffusive
(top) and localized branch (bottom). The marked maximum of $\xi(K)$ gives a precise determination of $K_c$. The parameters are:
$\varepsilon=0.4$ for a and b; $\varepsilon=0.8$ for c and d. $t$ varies up to $120$ kicks.} 
\end{figure}

\begin{table}
\begin{centering}
\begin{tabular}{|c|c|c|c|c|c|c|c|}
\hline 
$\varepsilon$ & $K_{1}-K_{2}$ & $K_{c}$ (exp) & $K_{c}$ (num) & $\ln \Lambda_{c}$(exp) & $\ln \Lambda_{c}$(num)\tabularnewline
\hline 
\hline 
0.2 & 7.0-14.0 & 8.85 $\pm$ 0.1 & 8.84  $\pm$ 0.47 & 2.1 $\pm$ 0.1 & 2.71 $\pm$ 0.44 \tabularnewline
\hline 
0.3 & 5.2-9.2 & 7.46 $\pm$ 0.05 & 7.71  $\pm$ 0.42 & 2.05 $\pm$ 0.08 & 2.22  $\pm$ 0.34 \tabularnewline
\hline 
0.4 & 4.5-8.5 & 6.75 $\pm$ 0.04 & 6.77  $\pm$ 0.52 & 1.95 $\pm$ 0.05 & 1.81 $\pm$ 0.47  \tabularnewline
\hline 
0.5 & 4.0-8.0 & 6.00 $\pm$ 0.04 & 5.93  $\pm$ 0.37 & 1.85 $\pm$ 0.05 & 1.36  $\pm$ 0.46 \tabularnewline
\hline 
0.6 & 3.4-7.4 & 5.59 $\pm$ 0.04 & 5.27  $\pm$ 0.35 & 1.75 $\pm$ 0.05 & 1.10  $\pm$ 0.30 \tabularnewline
\hline 
0.7 & 2.9-6.9 & 5.27 $\pm$ 0.03 & 4.99 $\pm$ 0.34 & 1.60 $\pm$ 0.05 & 0.94  $\pm$ 0.40 \tabularnewline
\hline 
0.8 & 2.8-6.8 & 4.84 $\pm$ 0.03 & 4.70 $\pm$ 0.43 & 1.52 $\pm$ 0.04 & 0.98  $\pm$ 0.31 \tabularnewline
\hline 
\end{tabular}
\par\end{centering}

\caption{\label{tab:Results} Experimental results on the determination
of the critical point of the metal-insulator Anderson transition,
for various values of the parameter $\varepsilon$ of the quasi-periodic kicked rotor. 
The second column indicates the range
of $K$ where data has been taken. The experimentally measured values
of both $K_{c}$ and $\Lambda_{c}$ are compared to the numerically
calculated values. 
The uncertainties on the experimental data are rather small.
The numerical data (with times up to $1000$ kicks) have similar uncertainties, but also
incorporate \textit{systematic} shifts of $(K_c, \Lambda_c)$ as a function of time 
\cite{Slevin:ScalingAnderson:PRL99} which cannot be estimated in the experimental data due to the restricted range of observation times $t \leq 120$ kicks.}
\end{table}

\begin{figure}
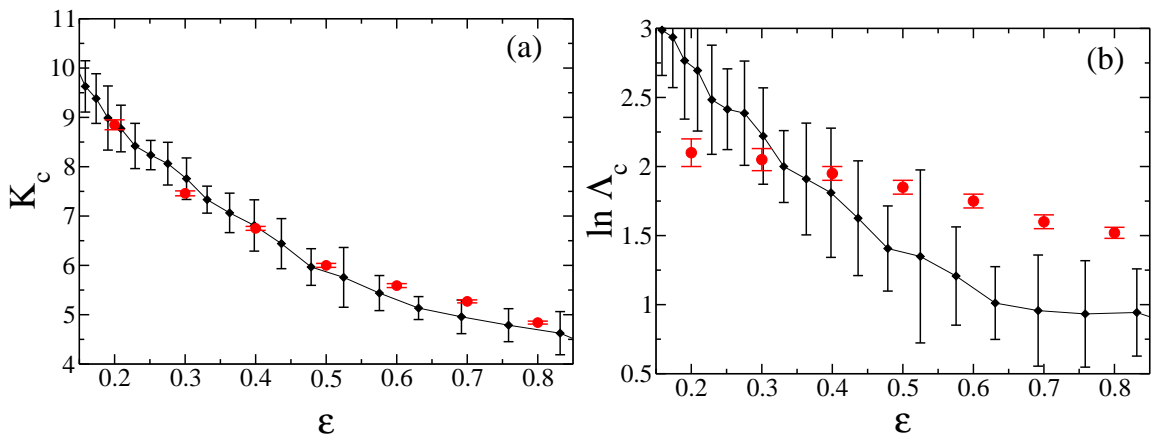

\begin{centering}
\includegraphics[width=0.48\linewidth]{figure3a.eps}
\includegraphics[width=0.48\linewidth]{figure3b.eps}
\par\end{centering}

\caption{\label{fig:Results}(a) Position of the critical point
$K_{c}(\varepsilon)$ , and (b) value of the critical $\Lambda_{c}(\varepsilon)$.
Numerical results (black diamonds) and experimental measurements (red circles) are represented with their associated error bars. The uncertainties 
on the experimental data are rather small, as can be directly seen in fig.~\ref{fig:Scaling}.
The numerical data (with times up to $1000$ kicks) have similar uncertainties, but also
incorporate systematic deviations of $(K_c, \Lambda_c)$ when estimated over various temporal ranges.
These systematic deviations cannot be easily measured in
the experiment (limited to $120$ kicks). 
In plot (a) one observes a very good agreement between
numerical and experimental results. The agreement is good in plot (b), except in the region of low $\varepsilon$ where
decoherence is expected to have a significant impact on the results and in the region of large $\varepsilon$ where the finite variance of the initial momentum distribution tends to increase $\Lambda_c$ at $120$ kicks 
but has only small effect on $K_c$.}
\end{figure}

In our experience, we measure
the population of the zero velocity class $\Pi(0;t)$ 
using Raman velocimetry \cite{Ringot:RamanSpectro:PRA01,Chabe:Polarization:OC07,Ringot:DiodeModule:EPJD99}.
This quantity is proportional to $\langle \pred^{2}(t)\rangle$, with a proportionality factor which depends
on the specific shape of $\Pi(\pred).$ This factor varies smoothly across the Anderson transition, so that the transition
can be studied using either   $\langle \pred^{2}(t)\rangle$ or $\Pi(0;t)$.
The scaling theory of localization~\cite{Abrahams:Scaling:PRL79,Lemarie:AndersonLong:PRA09}
predicts that $\left\langle \pred^{2}\right\rangle $ has characteristic
asymptotic behaviors in $t^{\alpha}$, with $\alpha=0$ in the localized
regime, $\alpha=2/3$ in the critical regime, and $\alpha=1$ in the
diffusive regime. This prediction has been fully confirmed by the 
experimental observations~\cite{Lemarie:CriticalStateAndersonTransition:PRL10}.
One can then define the \emph{scaling variable} \cite{Chabe:Anderson:PRL08,Lemarie:AndersonLong:PRA09}:
\begin{equation}
\Lambda(t) \equiv \frac{\left\langle \pred^{2}(t)\right\rangle }{t^{2/3}} \propto \frac{1}{\Pi(\pred=0;t)^{2}\ t^{2/3}}  \; .\label{eq:Lambdaexp}
\end{equation}
Asymptotically, $\Lambda(t) \propto t^{-2/3},t^0,t^{1/3}$ in the
localized, critical and diffusive regimes, respectively, so that  $\ln \Lambda(t)$ vs $\ln t^{1/3}$ displays a positive slope $1$ in the
diffusive regime, zero slope at the critical point and negative slope
$-2$ in the localized regime, which allows one to unambiguously identify
the critical point. However, experimental limitations prevent us from
performing measurements at large enough times (in our experiments
typically $t_{\mathrm{max}}=120$) to distinguish precisely between the localized and diffusive behaviors in the vicinity of the transition 
\footnote{Note however that for the parameters used here, $t_{\mathrm{max}}/t_{\mathrm{loc}}\sim10$
where $t_{\mathrm{loc}}$ is the localization time for the lowest $K$ value used
in each series at fixed $\varepsilon,$ so that thoe lowest point is clearly
in the localized regime.}. 
The main causes of this limitation is the falling (under gravity
action) of the cold atoms out of the standing wave and decoherence
induced by spontaneous emission. 

Fortunately, a technique known as
\emph{finite size scaling} (which is finite \emph{time} scaling in
our case), based on arguments derived from the so-called \emph{one
parameter scaling theory} of the Anderson transition \cite{Abrahams:Scaling:PRL79}
allows us to overcome this limitation. The application of this technique
to our problem has been discussed in details in previous works
\cite{Lemarie:AndersonLong:PRA09,Lemarie:These:09,Lemarie:UnivAnderson:EPL09};
let us just say here that it relies on the verified hypothesis that $\Lambda$ can be written as a one-parameter scaling function:
\begin{equation}\label{eq:scafunc}
 \Lambda = F\left(\frac{\xi(K)}{t^{1/3}}\right) \; ,
\end{equation}
with the scaling parameter $\xi(K)$ which plays the role of the localization length $p_\text{loc}$ in the localized regime and 
of the inverse of the diffusive constant in the diffusive regime. This method produces a rather precise determination
of the critical parameters $K_{c}(\varepsilon)$ and $\Lambda_c(\varepsilon)$ and of the critical exponent of the Anderson transition
\cite{Lopez:ExperimentalTestOfUniversality:PRL12} even from experimental
signals limited to a hundred of kicks or so. An example of such a
determination is presented in Fig.~\ref{fig:Scaling}. The critical
point corresponds to the tip at the right of the curve in Fig.~\ref{fig:Scaling}a and c,
at the intersection of the two clearly defined branches: A diffusive
(top) and a localized branch (bottom). By construction, in principle
$\xi(K)$ should diverge at the critical point, but the finite duration
of the experiment and decoherence effects produce a cutoff; however,
it still presents, as shown in Fig.~\ref{fig:Scaling}b and d a marked
maximum at the transition, and a careful fitting procedure taking
these effects into account \cite{Lopez:ExperimentalTestOfUniversality:PRL12}
allows a precise determination of $K_{c}$.
Once the value of $K_c$ has been determined according to the above technique, we measure the full momentum distribution, which
is found to be in excellent agreement with the predicted Airy shape, eq.~(\ref{eq:Airy}), as shown in \cite{Lemarie:CriticalStateAndersonTransition:PRL10}. A simple fit of the experimental data by an Airy function
allows to measure $\langle\pred ^2\rangle,$ hence $\Lambda_c$.

We have measured the value of the critical parameters $K_{c}(\varepsilon)$
and $\Lambda_c(\varepsilon)$ [Eq.~(\ref{eq:Lambdaexp})]
for a grid of 7 paths at constant $\varepsilon$ in the parameter
plane $(K,\varepsilon)$ (see Fig.~\ref{fig:Paths}) 
For each path, 50 uniformly spaced values of $K$ are used and the values of $\Pi(0;t)$
measured for each $K$ value; the initial and final values or $K$ are chosen symmetrically
with respect to the critical point. For each value of $K,$ an average
of 20 independent measurements is performed, a full path thus corresponds to
more than seven hours of data acquisition. Table~\ref{tab:Results}
gives the details of each path and the results obtained.

Figure~\ref{fig:Results} displays the experimental and numerical results. Plot (a) indicates
the position of the critical point $K_{c}(\varepsilon)$ and plot
(b) the critical value $\Lambda_{c}(\varepsilon)$. In both plots, experimental measurements are
indicated by red circles, numerical simulation results by black diamonds and are represented along with their error bars. The uncertainty of the numerical data
(see the following for a discussion of the numerical method) is evaluated from data up to $t=1000$ kicks and thus incorporates \textit{systematic} shifts of $(K_c, \Lambda_c)$ as a function of time 
\cite{Slevin:ScalingAnderson:PRL99} which cannot be estimated in the experimental data due to the restricted range of observation times $t \leq 120$ kicks. This results in larger numerical error bars than experimental ones.
Note also that a small uncertainty in $K_c$ implies a much larger error in $\Lambda_c$ due to its rapid variation as a function of $K$.    
In plot (a), one observes a very good agreement
between numerical and experimental results. In plot (b), the agreement
is good, except in the region
of low $\varepsilon$ which corresponds to high values of $K$ and
are thus more sensitive to decoherence effects. In the region of large $\varepsilon$, the finite variance of the initial momentum distribution tends to increase the experimental $\Lambda_c$, an effect which is not present
in the numerical data.

\section{Self-consistent theory of the anisotropy phase diagram}

We shall now try to describe theoretically the observed anisotropy dependences of the two critical parameters 
$K_c(\varepsilon)$ and $\Lambda_c(\varepsilon)$. The approach we shall follow is based on the self-consistent
theory of localization \cite{Vollhardt:SelfConsistentTheoryAnderson:92} which has been used successfully 
to predict numerous properties of the Anderson transition, and in particular the disorder vs energy \cite{Wolfle:theoPhaseDiagAndtransi:PRB90} 
and disorder vs anisotropy \cite{Soukoulis:AnisotropicAnderson:PRL96} phase diagrams of the 3D Anderson model. 
Moreover, the self-consistent theory of localization has been transposed to the case of the kicked rotor 
\cite{Altland:diagtheoKR:PRL93, Tian:diagtheoKR:PRB05}. We will use in the following a simple generalization 
of this latter approach
adapted to the case of the 3D anisotropic kicked pseudo-rotor \eqref{eqKR3DquasiperH} corresponding to the quasiperiodic 1D kicked rotor. 
 
The starting point is to consider the probability to go from a site $\boldsymbol \pred_i $ to a site $\boldsymbol \pred_f $ in $N$ steps, 
$P( \boldsymbol \pred_i , \boldsymbol \pred_f, t=N) \equiv \vert \langle  \boldsymbol \pred_f \vert U^N \vert \boldsymbol \pred_i \rangle\vert^2$. 
It consists in propagations mediated by the hopping 
amplitudes $\langle \boldsymbol \pred_{n+1} \vert J \vert \boldsymbol \pred_{n} \rangle$ and by collisions on 
the disorder represented by $ V(\boldsymbol{\pred})= e^{-i \phi_{\boldsymbol{\pred}}}$. Two important points are the following:
(i) one can consider in a first approximation the $\phi_{\boldsymbol{\pred}}$ as completely random 
phases \cite{Fishman:LocDynAnders:PRL82, Shepelyanky:Kq:PD87, Fishman:psudorandLoca:PRL88} and we will consider
quantities averaged over those phases, for example $\overline {P}( \boldsymbol \pred_i , \boldsymbol \pred_f, t=N)$ 
where the line over the quantity represents this averaging;
(ii) $\langle \boldsymbol \pred_{n+1} \vert J \vert \boldsymbol \pred_{n} \rangle$ plays the role of the 
disorder averaged Green's function (in the usual language of diagrammatic theory of transport in disordered systems 
\cite{akkermans2007mesoscopic}), that is the propagation between two scattering events. 
Indeed, when $\varepsilon=0$ and in the direction $p_1$, 
this is just a Bessel function which decreases exponentially fast for 
$\vert \boldsymbol \pred_{n+1} -   \boldsymbol \pred_{n} \vert \gg K/\kbar$ and one can thus 
see $K/\kbar$ as the analog of the mean free path, with the limit 
of small disorder corresponding to $K/\kbar \gg 1$.

One can attack the problem of the calculation of $\overline{P}$ by looking for propagation terms -- including of course interference patterns -- which survive the disorder averaging. 
At lowest order, 
the contribution containing no interference term to the 
probability $\overline{P}$ is called the Diffuson \cite{akkermans2007mesoscopic}. It corresponds to the classical
chaotic diffusion and can be shown to have a diffusive kernel
expressed in the reciprocal space $(\boldsymbol q , \omega)$ (conjugated to  $(\boldsymbol \pred , t)$) as \cite{Altland:diagtheoKR:PRL93, Tian:diagtheoKR:PRB05, Lemarie:These:09}:
\begin{equation}\label{eq:diffkern}
 \overline{P_D}(\boldsymbol q , \omega) =  \frac{1}{-i \omega + \sum_j D_{jj} q_j^2} \; .
\end{equation}
Here, the diffusive tensor $\boldsymbol D$ -- computed in \cite{Lemarie:KR3DClassic:JMO10} for large $K$  -- is anisotropic, but diagonal, with: 
\begin{equation}
 \eqalign{D_{11} =  \frac{K^2}{4 \kbar^2} \left(1+ \frac{\varepsilon^2}{4}\right) \; ,\cr 
D_{22} = D_{33}  =\frac{K^2 \varepsilon^2}{16 \, \kbar^2} \; .}\label{eq:diffcoeff}
\end{equation}
This anisotropic diffusive kernel is valid at long times and on large scale in momentum space, that is in the so-called
hydrodynamic limit $\omega \ll 1$ and $q_j k_j \ll 1$, 
with $k_j$ the mean free path along direction $j$ which is such that $D_{jj} = k_j^2/4$. 
Equation \eqref{eq:diffkern} means that in the regime of long times and large distances (in momentum space), 
we should have a diffusive transport with $\langle \pred_j^2 \rangle= 2 D_{jj} t$. 
This is certainly not the case near the Anderson transition, which implies
that we must go beyond the Diffuson approximation.

The simplest interferential correction -- known as weak localization correction -- is due to the
constructive interference between pairs of time-reversed paths
\footnote{The quasi-periodic kicked rotor or the equivalent periodic 3D pseudo-rotor is not invariant
by time reversal. However, the Hamiltonian, eq.~(\ref{eqKR3DquasiperH}), is invariant by the product
of time reversal and parity. The existence of such an anti-unitary symmetry is sufficient
for the system to belong to the Orthogonal universality class, and consequently for the existence 
of the weak localization correction.}, or equivalently to the maximally crossed diagrams, or Cooperon. 
The net effect of these interferential contributions is to increase the return probability at the initial point
and to decrease the diffusion constant. It is possible to compute perturbatively the weak localization
correction as an integral (see below) depending on the diffusion constant itself.  
Contributions from higher orders are extremely complicated and
there is no systematic way of summing them all. 

The self-consistent theory of localization
is a simple attempt at approximately summing the most important contributions: instead of computing
the weak localization correction using the raw diffusion constant, one uses the one renormalized
because of weak localization. The whole thing must of course be self-consistent, so that the diffusion
constant computed taking into account the weak localization correction is equal to the one
input in the calculation of this correction. The price to pay is that one can no longer define
a single diffusion constant -- or, in the anistropic case, a single diffusion tensor -- but must
introduce a frequency-dependent diffusion constant (or diffusion tensor). This is nevertheless quite natural
if one wants to describe the transition from a diffusive behaviour as short time (large frequency)
when interference terms are small to a localized behaviour at long time (small frequency).
The intensity propagator $ \overline{P}$ takes then the approximate form:
$ \overline{P}(\boldsymbol q , \omega) =  \frac{1}{-i \omega + \sum_j {\cal D}_{jj}(\omega) q_j^2} $
with the frequency dependent diffusion constant following the self-consistent equation \cite{Lemarie:These:09,Vollhardt:SelfConsistentTheoryAnderson:92, Altland:diagtheoKR:PRL93}:
\begin{eqnarray}
{\cal D}_{ii}(\omega) = D_{ii} - 2 {\cal D}_{ii}(\omega) \int \frac{d^3 \boldsymbol q}{(2 \pi)^3} 
\dfrac{1}{-i \omega + \sum_j {\cal D}_{jj}(\omega) q_j^2} \; .
\label{eq:self-cons}
\end{eqnarray}

Quite remarkably, this approach is able to account for a certain number of observed features: it predicts a transition 
between a metallic phase of diffusive transport for $K>K_c(\varepsilon)$ where
${\cal D}_{ii}(\omega) \underset{\omega \rightarrow 0}{\sim} {\cal D}_{ii}(0)>0$, to a localized phase for 
$K<K_c(\varepsilon)$
where ${\cal D}_{ii}(\omega)\underset{\omega \rightarrow 0}{\sim} -i \omega \; {\pred_\text{loc}}_i^2$ with 
${\pred_\text{loc}}_i$ the localization length along direction $i$. At the threshold,
the transport is predicted to follow an anomalous diffusion with ${\cal D}_{ii}(\omega) \sim (-i \omega)^{1/3}$ 
and this implies the Airy shape of the critical state observed experimentally 
\cite{Lemarie:CriticalStateAndersonTransition:PRL10}. In the following, we will
calculate explicitly the critical parameters $K_c(\varepsilon)$ and 
$\Lambda_c(\varepsilon)$ from \eqref{eq:self-cons} and show that it 
also complies well with the experimental observations.

One shall evaluate the integral on the right hand side of 
equation \eqref{eq:self-cons}. It is important to remark that,
although the system is anistropic and thus 3 different equations (\ref{eq:self-cons}) have
to be solved simultaneously, they in fact follow exactly the same renormalization scheme: dividing
eq.~(\ref{eq:self-cons}) by $D_{ii}$ produces the same equation in all 3 dimensions. In other words,
there is no anomaly in the anisotropic character at the critical point.
 
It is well known \cite{Vollhardt:SelfConsistentTheoryAnderson:92} that in 
dimension $d \geq 2$, the results
of the self-consistent theory are cutoff dependent. Indeed, the integral in \eqref{eq:self-cons}
diverges at large $q$ and must be limited  to 
$q_j < q_j^{\text{max}}$, where $q_j^{\text{max}}$ is a
cutoff on the order of $k_j^{-1},$ i.e. the inverse of the mean free path.
In the following, we will take \cite{Soukoulis:AnisotropicAnderson:PRL96} 
$q_j^{\text{max}} \equiv C_1 k_j^{-1}  = C_1/(2\sqrt{D_{jj}})$ with $C_1$ a numerical constant
of the order of one.
We make the following change of variables: $Y_j = \sqrt{\frac{{\cal D}_{jj} (\omega)}{-i \omega} } q_j$ 
and define the rescaled cutoff: 
$\ell(\omega) \equiv  \sqrt{\frac{{\cal D}_{jj} (\omega)}{-i \omega} } q_j^{\text{max}}$ (from Eq. \eqref{eq:self-cons} 
it is clear that the ratio ${\cal D}_{jj}(\omega)/D_{jj}$ is isotropic, thus $\ell(\omega)$ is isotropic). One obtains:
\begin{eqnarray}
\frac{{\cal D}_{ii}(\omega)}{D_{ii}} = 1 - \frac{C_1}{2\pi^2} \frac{1}{\sqrt{D_{11}D_{22}D_{33}}} \left[1 - \frac{\tan^{-1}\ell(\omega)}{\ell(\omega)}\right] \; .
\label{eq:crit-self-cons}
\end{eqnarray}

The threshold $K_c(\varepsilon)$ of the Anderson transition is then approached 
from the diffusive regime, which is caracterized by 
$\frac{{\cal D}_{ii}(\omega)}{D_{ii}} \underset{\omega \rightarrow 0}{\rightarrow} \frac{{\cal D}_{ii}(0)}{D_{ii}}
\underset{K \rightarrow K_c}{\rightarrow} 0$ and 
$\ell(\omega) \underset{\omega \rightarrow 0}{\rightarrow}   \infty$. 
Therefore, $K_c(\varepsilon)$ is such that:
\begin{equation}
 D_{11} D_{22} D_{33} = \frac{{C_1}^2}{4 \pi^4} \; . \label{eq:threshold}
\end{equation}
From the above expressions \eqref{eq:diffcoeff} for the diffusion tensor, 
one deduces the following dependence of the threshold vs anisotropy:
\begin{equation}\label{eq:Kcvsepssimple}
 K_c(\varepsilon) = \left(\frac{2^4 C_1 }{\pi^{2}}\right)^{1/3} \dfrac{\kbar}{(\varepsilon^2 \sqrt{1+\varepsilon^2/4})^{1/3}} \; .
\end{equation}

The self-consistent theory allows also for a determination of $\Lambda_c(\varepsilon)$. 
In fact, at finite but sufficiently small $\omega$ (i.e. at sufficiently large times), $\ell(\omega)$ is large 
and one can evaluate the right hand side of \eqref{eq:crit-self-cons} at 
the lowest order in $1/\ell(\omega)$ 
which gives: 
\begin{equation}
\frac{{\cal D}_{ii}(\omega)}{(-i\omega)^{1/3}} = \frac{1}{(2\pi)^{2/3}} \frac{D_{11}}{(D_{11}D_{22}D_{33})^{1/3}} \; .
\end{equation}
We know from the study of the critical state of the Anderson 
transition \cite{Lemarie:CriticalStateAndersonTransition:PRL10} that 
$\frac{{\cal D}_{11}(\omega)}{(-i \omega)^{1/3}} = \frac{\Gamma(2/3)}{3} \Lambda_c$ which
allows us to write:
\begin{equation}
 \Lambda_c(\varepsilon) = \frac{3}{\Gamma(2/3)} \left(\frac{1}{2 \pi}\right)^{2/3} \left(\frac{D_{11}^2}{D_{22} D_{33}}\right)^{1/3} \; . \label{eq:lambdac}
\end{equation}
Using the diffusion tensor relations \eqref{eq:diffcoeff}, one obtains finally:
\begin{equation}
\Lambda_{c}(\varepsilon)=\frac{3}{\Gamma(2/3) }  
\left(\frac{2}{\pi}\right)^{2/3}\left(\dfrac{1+\varepsilon^{2}/4}{\varepsilon^{2}}\right)^{2/3} \; .\label{eq:Lambdacth}
\end{equation}

Equations \eqref{eq:Kcvsepssimple} and \eqref{eq:Lambdacth} are the most important results of this section. They predict that the threshold $K_c(\varepsilon)$ 
and the critical anomalous diffusion parameter $\Lambda_c(\varepsilon)$ diverge at large anisotropy as $K_c(\varepsilon)\sim \varepsilon^{-2/3}$ and $\Lambda_c(\varepsilon)\sim \varepsilon^{-4/3}$. 
Therefore, when $\varepsilon=0$ we recover the case of the 1D periodic kicked rotor which is always localized whatever the kicking amplitude $K$.

\section{Experimental and numerical tests of the self-consistent theory of the anisotropy phase diagram}

\begin{figure}
\begin{centering}
\includegraphics[width=0.6\linewidth]{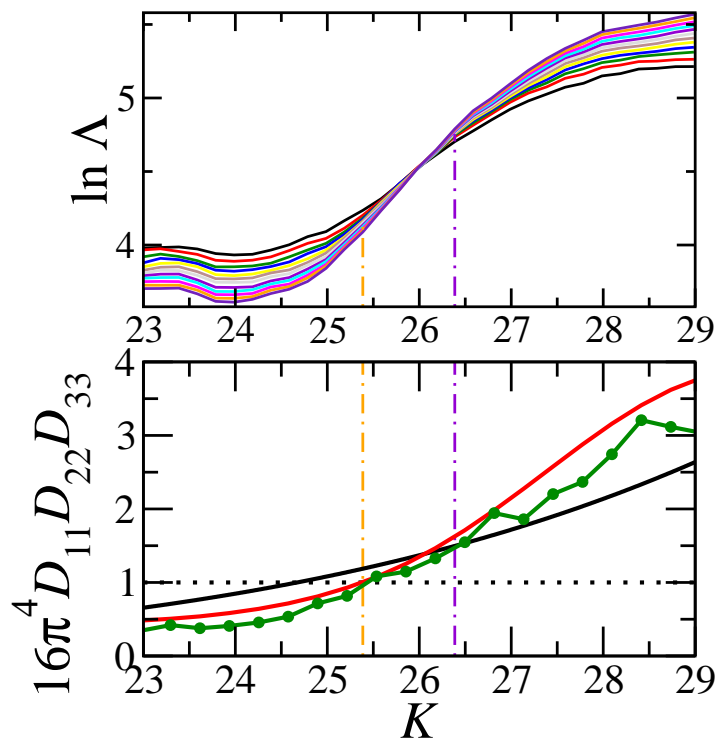}
\end{centering}
\caption{\label{fig:method-threshold} 
Method of determination of the threshold $K_c(\varepsilon)$. 
Upper panel: Numerical data for $\ln \Lambda = \ln \frac{\langle \pred^2 \rangle}{t^{2/3}}$ vs $K$ 
at different times ranging from $t=36$ to $956$. The threshold corresponds to the crossing of these 
lines where we have the critical anomalous diffusion $ \langle \pred^2 \rangle \sim t^{2/3}$, 
whereas for $K<K_c(\varepsilon)$, $\Lambda(t)$ decreases at long time (localized regime) and 
for $K>K_c(\varepsilon)$ $\Lambda(t)$ increases at long time (metallic regime). 
The uncertainty region (between the orange and violet dash-dotted lines) corresponds to the region 
where the evolution of $\Lambda(t)$ is not monotonous.
Lower panel: The different degrees of approximation for $16 \pi^4 D_{11} D_{22} D_{33}$. 
According to Eq.\ \eqref{eq:threshold} (with $C_1=1/2$, see text),
this quantity should be equal to unity at the critical point.
The black line corresponds to the simple analytic prediction \eqref{eq:diffcoeff} for the diffusion tensor. 
The red line shows the theoretical prediction incorporating oscillating corrections 
for the diffusion tensor (see Eq.\ \eqref{eq:oscil} and text). The green line with points shows numerical data for the 
diffusion tensor $\boldsymbol D$ of the 3D kicked rotor \eqref{eqKR3DquasiperH}, at short time. 
The parameters are: $\kbar=2.89$, $\omega_2 = 2 \pi \sqrt{5}$ and 
$\omega_3= 2 \pi \sqrt{13}$, $\varepsilon = 0.036$. 
}
\end{figure}

\begin{figure}
\begin{centering}
\includegraphics[clip,width=0.6\linewidth]{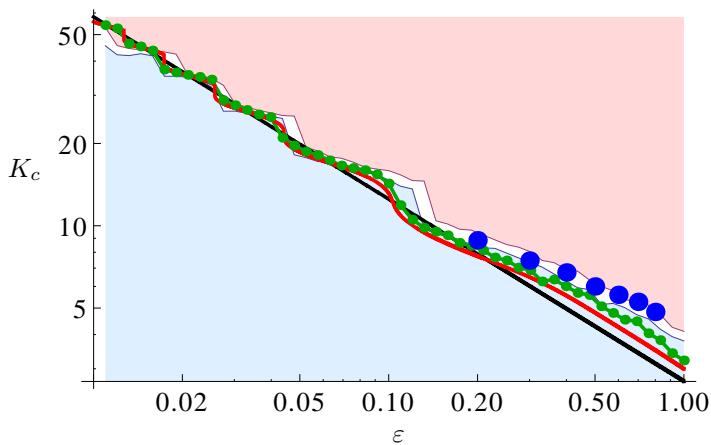}
\par\end{centering}
\caption{\label{fig:threshold} Threshold of the Anderson transition vs anisotropy (log-log scale). 
The anisotropy dependence of $K_c(\varepsilon)$ with the associated uncertainty is determined 
from numerical simulations of the dynamics of the quasi-periodic kicked
rotor and is represented by the white filled region between the blue (localized) and pink (metallic) filled regions. 
The three degrees of approximation
of the self-consistent theory prediction \eqref{eq:threshold} (with $C_1=1/2$) are shown: (i) the black line corresponds to the simple analytic prediction
\eqref{eq:Kcvsepssimple}, (ii) the red line incorporates oscillating 
corrections for the diffusion tensor (see Eq.~\eqref{eq:oscil} and text) while (iii) the green line with points corresponds
to numerical data for 
the diffusion tensor $\boldsymbol D$ of the 3D kicked rotor \eqref{eqKR3DquasiperH}, at short time.
The blue points represent the experimental data. 
The parameters are: $\kbar=2.89$, $\omega_2 = 2 \pi \sqrt{5}$ and $\omega_3= 2 \pi \sqrt{13}$. }
\end{figure}

\begin{figure}
\begin{centering}
\includegraphics[clip,width=0.6\linewidth]{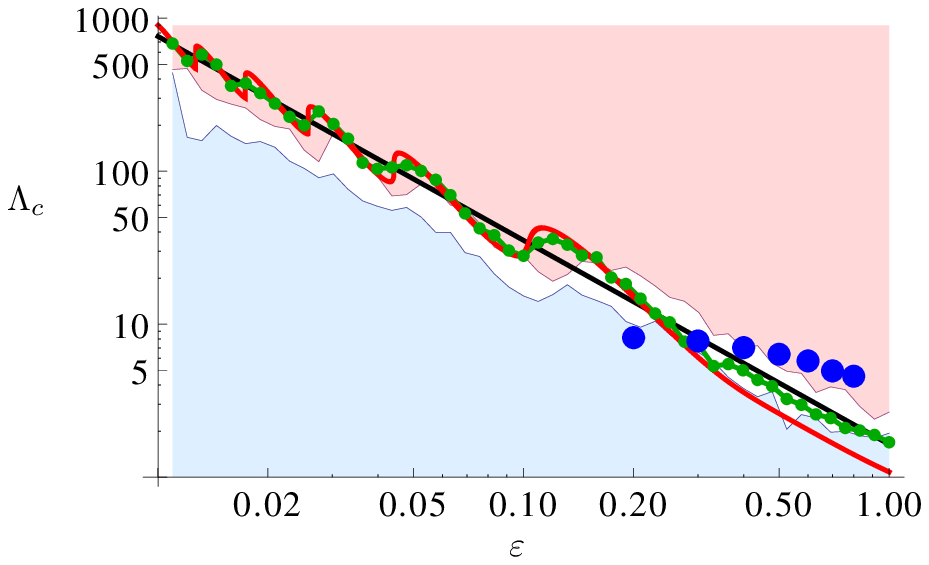}
\par\end{centering}
\caption{\label{fig:lambdac} Critical parameter $\Lambda_c$ vs anisotropy (log-log scale). 
The anisotropy dependence of $\Lambda_c(\varepsilon)$ with the associated uncertainty is 
determined from numerical simulations of the dynamics of the quasi-periodic kicked
rotor and is represented by the white filled region between the blue (localized) and 
pink (metallic) filled regions. The prediction \eqref{eq:lambdac}
of the self-consistent theory is shown with the three different degrees of approximations considered: (i) the simple analytic prediction \eqref{eq:Lambdacth} is shown in black line, 
(ii) the red lines is based on the analytic prediction Eq.\ \eqref{eq:oscil} for the diffusion tensor incorporating oscillating corrections in $K$ and $\kbar$ and 
(iii) numerical data for the diffusion 
tensor $\boldsymbol D$ of the 3D kicked rotor at short time \eqref{eqKR3DquasiperH} give the green line with points.
The blue points represent the experimental data. The parameters are: $\kbar=2.89$, $\omega_2 = 2 \pi \sqrt{5}$ and $\omega_3= 2 \pi \sqrt{13}$.}
\end{figure}

In order to test the predictions of the self-consistent theory,
we have performed numerical simulations of the dynamics of the quasi-periodic 
kicked rotor \eqref{eq:Hqpkr}. We have determined the critical parameters 
$K_c(\varepsilon)$ and $\Lambda_c(\varepsilon)$ from the crossing of the curves 
$\ln \Lambda = \ln \frac{\langle \pred^2 \rangle}{t^{2/3}}$ vs $K$ at different times 
(see figure \ref{fig:method-threshold}). At the critical point, $\Lambda(t)$ is a 
constant $\Lambda(t)=\Lambda_c(\varepsilon)$ corresponding to the critical 
anomalous diffusion $ \langle \pred^2 \rangle \sim t^{2/3}$ and the crossing of the curves in fig.\ \ref{fig:method-threshold}, whereas for 
$K<K_c(\varepsilon)$ ($K>K_c(\varepsilon)$, resp.), $\Lambda(t)$ decreases (increases, resp.) as time increases. We have
evaluated the uncertainty of the parameters $(K_c, \Lambda_c)$ by the region 
where the evolution of $\Lambda(t)$ is not monotonous (due to systematic shifts of $(K_c,\Lambda_c)$ as a function of time \cite{Slevin:ScalingAnderson:PRL99}).
The results are represented in fig.\ \ref{fig:threshold} for $K_c(\varepsilon)$ and in fig.\ \ref{fig:lambdac} for $\Lambda_c(\varepsilon)$ 
by the white filled region between the blue filled region where the system is localized and pink filled region where the dynamics is diffusive. The data seem to follow an algebraic increase as the 
anisotropy parameter $\varepsilon$ decreases (linear dependence in log-log scale as in figs.\ \ref{fig:threshold} and \ref{fig:lambdac}), on top of which oscillations are also clearly seen. 

The self-consistent theory discussed in the previous section predicts that the critical regime of the Anderson transition
is given by Eq.~(\ref{eq:threshold}). Various approximations -- leading to slightly different predictions for the
position of the critical point -- can be used for the values of the components of the diffusion tensor:
\begin{itemize}
\item (i) The simplest approximation is to use eq.~\eqref{eq:diffcoeff}, valid asymptotically for large $K.$ 
This results in the simple analytic predictions \eqref{eq:Kcvsepssimple} and \eqref{eq:Lambdacth},  represented by black lines in figures
\ref{fig:threshold} and \ref{fig:lambdac}. The algebraic dependences of $K_c(\varepsilon)$ and $\Lambda_c(\varepsilon)$ are well accounted for by these simple predictions, however they fail to reproduce the oscillating corrections observed in the numerical data.

\item (ii) The theoretical prediction \eqref{eq:diffcoeff} for the 
diffusion tensor $\boldsymbol D$ miss the oscillations of the diffusion 
tensor of the 3D kicked pseudo-rotor \eqref{eqKR3DquasiperH} vs $K$ and $\kbar$. Such oscillations are well known in the case of the 1D periodic kicked 
rotor \cite{Rechester:KRDiffCoeff:PRA81, Shepelyanky:Kq:PD87} and arise due to subtle temporal correlation effects. In our case, we have checked that they could be described approximately by the known oscillating 
form \cite{Shepelyanky:Kq:PD87}, but only along direction 1:
\begin{equation}\label{eq:oscil}
\eqalign{
 \tilde{D}_{11} \approx D_{11} \times  \lbrace 1 - 2 J_2(\tilde{K}) [1 - J_2(\tilde{K})] \rbrace \; ,\cr 
 \tilde{D}_{22}=  \tilde{D}_{33} = D_{22} = D_{33} \; .}
\end{equation}
with  $\tilde{K} \equiv K \frac{\sin \kbar/2}{\kbar/2}$ and $J_2$ the usual Bessel function. The use of 
the above equation for the diffusion tensor allows for a better analytical description of the anisotropy phase diagram, see the 
red lines in figures \ref{fig:threshold} and \ref{fig:lambdac}. 

\item (iii) The third type of approximation relies on a direct numerical calculation of the diffusion tensor $\boldsymbol D$ of the  
3D kicked rotor \eqref{eqKR3DquasiperH} at short time by
a linear fitting procedure over the first ten kicks of 
$\langle \pred_i^2 \rangle = 2 D_{ii} t$. This gives a numerical prediction of the self-consistent theory for the anisotropy phase diagram 
represented by the  green lines with points in figures \ref{fig:threshold} and \ref{fig:lambdac}. 
They clearly show oscillations around the power law behaviors of 
$K_c(\varepsilon)$ and $\Lambda_c(\varepsilon)$, and are in very good agreement with the numerical data.
\end{itemize}

Last but not least, we clearly see in fig.\ \ref{fig:threshold} and \ref{fig:lambdac} that the predictions of the self-consistent theory agree also very well with the experimental data represented by blue points. 
Therefore, the self-consistent theory appears to be a powerful way to describe the Anderson transition with the quasi-periodic kicked rotor.

\section{Conclusion}

In conclusion, we presented in this work a rather complete study of
the anisotropy phase diagram of
the Anderson transition in the quasiperiodic kicked rotor. Numerical
and experimental results were found to be in good agreement with each
other, and theoretical expressions based on the self-consistent theory
of the Anderson transition correctly describe the shape dependence
of these functions. These results bring an important additional brick
to our understanding of the Anderson transition, and put in an even
firmer ground the status of the quasiperiodic kicked rotor as one of the simplest
-- if not the simplest -- 
cold atom system for experimentally studying Anderson localization.

\ack{
Laboratoire de Physique des Lasers, Atomes et Mol\'ecules is Unit\'e
Mixte de Recherche 8523 of CNRS. Work partially supported by Agence
Nationale de la Recherche (LAKRIDI grant) and ``Labex'' CEMPI.
CPU time on various computers has been provided by GENCI.}

\section*{References}
\bibliographystyle{unsrt}
\bibliography{ArtDataBase}

\begin{thebibliography}{10}

\bibitem{Anderson:LocAnderson:PR58}
P.~W. Anderson.
\newblock {Absence of Diffusion in Certain Random Lattices}.
\newblock {\em Phys. Rev.}, 109(5):1492--1505, 1958.

\bibitem{Billy:AndersonBEC1D:N08}
J.~Billy, V.~Josse, Z.~Zuo, A.~Bernard, B.~Hambrecht, P.~Lugan, D.~Cl{\'e}ment,
  L.~Sanchez-Palencia, P.~Bouyer, and A.~Aspect.
\newblock {Direct observation of Anderson localization of matter-waves in a
  controlled disorder}.
\newblock {\em Nature (London)}, 453:891--894, 2008.

\bibitem{Roati:AubryAndreBEC1D:N08}
G.~Roati, C.~d'Errico, L.~Fallani, M.~Fattori, C.~Fort, M.~Zaccanti,
  G.~Modugno, M.~Modugno, and M.~Inguscio.
\newblock {Anderson localization of a non-interacting Bose-Einstein
  condensate}.
\newblock {\em Nature (London)}, 453:895--898, 2008.

\bibitem{Kondov:ThreeDimensionalAnderson:S11}
S.~S. Kondov, W.~R. McGehee, J.~J. Zirbel, and B.~DeMarco.
\newblock {Three-Dimensional Anderson Localization of Ultracold Matter}.
\newblock {\em Science}, 334(6052):66 --68, 2011.

\bibitem{Jendrzejewski:AndersonLoc3D:NP12}
F.~Jendrzejewski, A.~Bernard, K.~Mueller, P.~Cheinet, V.~Josse, M.~Piraud,
  L.~Pezz{\'e}, L.~Sanchez-Palencia, A.~Aspect, and P.~Bouyer.
\newblock {Three-dimensional localization of ultracold atoms in an optical
  disordered potential}.
\newblock {\em Nature Physics}, 8:398, 2012.

\bibitem{Efetov:SupersymmetryInDisorder:97}
K.~Efetov.
\newblock {\em {Supersymmetry in Disorder and Chaos}}.
\newblock {Cambridge University Press}, {Cambridge, UK}, 1997.

\bibitem{Casati:IncommFreqsQKR:PRL89}
G.~Casati, I.~Guarneri, and D.~L. Shepelyansky.
\newblock {Anderson transition in a one-dimensional system with three
  incommensurable frequencies}.
\newblock {\em Phys. Rev. Lett.}, 62(4):345--348, 1989.

\bibitem{Casati:LocDynFirst:LNP79}
G.~Casati, B.~V. Chirikov, J.~Ford, and F.~M. Izrailev.
\newblock {\em {Stochastic behavior of a quantum pendulum under periodic
  perturbation}}, volume~93, pages 334--352.
\newblock {Springer-Verlag}, {Berlin, Germany}, 1979.

\bibitem{Chabe:Anderson:PRL08}
J.~Chab{\'e}, G.~Lemari{\'e}, B.~Gr{\'e}maud, D.~Delande, P.~Szriftgiser, and
  J.~C. Garreau.
\newblock {Experimental Observation of the Anderson Metal-Insulator Transition
  with Atomic Matter Waves}.
\newblock {\em Phys. Rev. Lett.}, 101(25):255702, 2008.

\bibitem{Lemarie:AndersonLong:PRA09}
G.~Lemari{\'e}, J.~Chab{\'e}, P.~Szriftgiser, J.~C. Garreau, B.~Gr{\'e}maud,
  and D.~Delande.
\newblock {Observation of the Anderson metal-insulator transition with atomic
  matter waves: Theory and experiment}.
\newblock {\em Phys. Rev. A}, 80(4):043626, 2009.

\bibitem{Lemarie:CriticalStateAndersonTransition:PRL10}
G.~Lemari{\'e}, H.~Lignier, D.~Delande, P.~Szriftgiser, and J.~C. Garreau.
\newblock {Critical State of the Anderson Transition: Between a Metal and an
  Insulator}.
\newblock {\em Phys. Rev. Lett.}, 105(9):090601, 2010.

\bibitem{Lopez:ExperimentalTestOfUniversality:PRL12}
M.~Lopez, J.-F. Cl{\'e}ment, P.~Szriftgiser, J.~C. Garreau, and D.~Delande.
\newblock {Experimental Test of Universality of the Anderson Transition}.
\newblock {\em Phys. Rev. Lett.}, 108(9):095701, 2012.

\bibitem{Ringot:Bicolor:PRL00}
J.~Ringot, P.~Szriftgiser, J.~C. Garreau, and D.~Delande.
\newblock {Experimental Evidence of Dynamical Localization and Delocalization
  in a Quasiperiodic Driven System}.
\newblock {\em Phys. Rev. Lett.}, 85(13):2741--2744, 2000.

\bibitem{Lignier:SubFMecs:EPL05}
H.~Lignier, J.~C. Garreau, P.~Szriftgiser, and D.~Delande.
\newblock {Quantum diffusion in the quasiperiodic kicked rotor}.
\newblock {\em EPL (Europhysics Letters)}, 69(3):327--333, 2005.

\bibitem{Lignier:Reversibility:PRL05}
H.~Lignier, J.~Chab{\'e}, D.~Delande, J.~C. Garreau, and P.~Szriftgiser.
\newblock {Reversible Destruction of Dynamical Localization}.
\newblock {\em Phys. Rev. Lett.}, 95(23):234101, 2005.

\bibitem{Chirikov:ChaosClassKR:PhysRep79}
B.~V. Chirikov.
\newblock {A universal instability of many-dimensional oscillator systems}.
\newblock {\em Phys. Rep.}, 52(5):263--379, 1979.

\bibitem{Fishman:LocDynAnders:PRL82}
S.~Fishman, D.~R. Grempel, and R.~E. Prange.
\newblock {Chaos, Quantum Recurrences, and Anderson Localization}.
\newblock {\em Phys. Rev. Lett.}, 49(8):509--512, 1982.

\bibitem{Altland:TheoryKR:PRL96}
Alexander Altland and Martin~R. Zirnbauer.
\newblock Field theory of the quantum kicked rotor.
\newblock {\em Phys. Rev. Lett.}, 77:4536--4539, Nov 1996.

\bibitem{Tian:TheoryKR:NJP10}
Chushun Tian and Alexander Altland.
\newblock Theory of localization and resonance phenomena in the quantum kicked
  rotor.
\newblock {\em New Journal of Physics}, 12(4):043043, 2010.

\bibitem{Lemarie:These:09}
G.~Lemari{\'e}.
\newblock {\em {Transition d'Anderson avec des ondes de mati{\`e}re
  atomiques}}.
\newblock PhD thesis, {Universit{\'e} Pierre et Marie Curie}, {Paris}, 2009.

\bibitem{Tian:TheoryAndersonTransition:PRL11}
C.~Tian, A.~Altland, and M.~Garst.
\newblock {Theory of the Anderson Transition in the Quasiperiodic Kicked
  Rotor}.
\newblock {\em Phys. Rev. Lett.}, 107(7):074101, 2011.

\bibitem{Soukoulis:AnisotropicAnderson:PRL96}
I.~Zambetaki, Qiming Li, E.~N. Economou, and C.~M. Soukoulis.
\newblock Localization in highly anisotropic systems.
\newblock {\em Phys. Rev. Lett.}, 76:3614--3617, May 1996.

\bibitem{Slevin:ScalingAnderson:PRL99}
K.~Slevin and T.~Ohtsuki.
\newblock {Corrections to Scaling at the Anderson Transition}.
\newblock {\em Phys. Rev. Lett.}, 82(2):382--385, 1999.

\bibitem{Ringot:RamanSpectro:PRA01}
J.~Ringot, P.~Szriftgiser, and J.~C. Garreau.
\newblock {Subrecoil Raman spectroscopy of cold cesium atoms}.
\newblock {\em Phys. Rev. A}, 65(1):013403, 2001.

\bibitem{Chabe:Polarization:OC07}
J.~Chab{\'e}, H.~Lignier, P.~Szriftgiser, and J.~C. Garreau.
\newblock {Improving Raman velocimetry of laser-cooled cesium atoms by
  spin-polarization}.
\newblock {\em Opt. Commun.}, 274:254--259, 2007.

\bibitem{Ringot:DiodeModule:EPJD99}
J.~Ringot, Y.~Lecoq, J.~Garreau, and P.~Szriftgiser.
\newblock {Generation of phase-coherent laser beams for Raman spectroscopy and
  cooling by direct current modulation of a diode laser}.
\newblock {\em Eur. Phys. J. D}, 7(3):285--288, 1999.

\bibitem{Abrahams:Scaling:PRL79}
E.~Abrahams, P.~W. Anderson, D.~C. Licciardello, and T.~V. Ramakrishnan.
\newblock {Scaling Theory of Localization: Absence of Quantum Diffusion in Two
  Dimensions}.
\newblock {\em Phys. Rev. Lett.}, 42(10):673--676, 1979.

\bibitem{Lemarie:UnivAnderson:EPL09}
G.~Lemari{\'e}, B.~Gr{\'e}maud, and D.~Delande.
\newblock {Universality of the Anderson transition with the quasiperiodic
  kicked rotor}.
\newblock {\em EPL (Europhysics Letters)}, 87:37007, 2009.

\bibitem{Vollhardt:SelfConsistentTheoryAnderson:92}
D.~Vollhardt and P.~W{\"o}lfle.
\newblock {Self-Consistent Theory of Anderson Localization}.
\newblock In {Hanke, W. and Kopaev Yu. V.}, editor, {\em {Electronic Phase
  Transitions}}, pages 1--78. {Elsevier}, 1992.

\bibitem{Wolfle:theoPhaseDiagAndtransi:PRB90}
J.~Kroha, T.~Kopp, and P.~W\"olfle.
\newblock Self-consistent theory of anderson localization for the tight-binding
  model with site-diagonal disorder.
\newblock {\em Phys. Rev. B}, 41:888--891, Jan 1990.

\bibitem{Altland:diagtheoKR:PRL93}
Alexander Altland.
\newblock Diagrammatic approach to anderson localization in the quantum kicked
  rotator.
\newblock {\em Phys. Rev. Lett.}, 71:69--72, Jul 1993.

\bibitem{Tian:diagtheoKR:PRB05}
C.~Tian, A.~Kamenev, and A.~Larkin.
\newblock Ehrenfest time in the weak dynamical localization.
\newblock {\em Phys. Rev. B}, 72:045108, Jul 2005.

\bibitem{Shepelyanky:Kq:PD87}
D.~L. Shepelyansky.
\newblock {Localization of diffusive excitation in multi-level systems}.
\newblock {\em Physica D}, 28:103--114, 1987.

\bibitem{Fishman:psudorandLoca:PRL88}
Meir Griniasty and Shmuel Fishman.
\newblock Localization by pseudorandom potentials in one dimension.
\newblock {\em Phys. Rev. Lett.}, 60:1334--1337, Mar 1988.

\bibitem{akkermans2007mesoscopic}
E.~Akkermans and G.~Montambaux.
\newblock {\em Mesoscopic physics of electrons and photons}.
\newblock Cambridge Univ Pr, 2007.

\bibitem{Lemarie:KR3DClassic:JMO10}
G.~Lemari{\'e}, D.~Delande, J.~C. Garreau, and P.~Szriftgiser.
\newblock {Classical diffusive dynamics for the quasiperiodic kicked rotor}.
\newblock {\em J. Mod. Opt.}, 57(19):1922--1927, 2010.

\bibitem{Rechester:KRDiffCoeff:PRA81}
A.~B. Rechester, M.~N. Rosenbluth, and R.~B. White.
\newblock {Fourier-space paths applied to the calculation of diffusion for the
  Chirikov-Taylor model}.
\newblock {\em Phys. Rev. A}, 23(5):2664--2672, 1981.

\end{thebibliography}

\end{document}